\documentclass[twocolumn,showpacs,preprintnumbers,amsmath,amssymb]{revtex4}
\usepackage{graphicx}% Include figure files
\usepackage{dcolumn}% Align table columns on decimal point
\usepackage{bm}% bold math

\begin{document}
\newcommand{\be}{\begin{equation}}   \newcommand{\ee}{\end{equation}}
\newcommand{\mean}[1]{\left\langle #1 \right\rangle}
\newcommand{\abs}[1]{\left| #1 \right|}
\newcommand{\set}[1]{\left\{ #1 \right\}}
\newcommand{\la}{\langle}
\newcommand{\ra}{\rangle}
\newcommand{\lb}{\left(}
\newcommand{\rb}{\right)}
\newcommand{\norm}[1]{\left\|#1\right\|}
\newcommand{\RA}{\rightarrow}
\newcommand{\tet}{\vartheta}
\newcommand{\eps}{\varepsilon}
\newcommand{\tNN}{\tilde{\mathbf{X}}_n^{NN}}
\newcommand{\NN}{\mathbf{X}_n^{NN}}
\newcommand{\ber}{\begin{eqnarray}}
\newcommand{\eer}{\end{eqnarray}}

\preprint{APS/123-QED}

\title{Noise reduction for flows using nonlinear constraints}

\author{Krzysztof Urbanowicz}
 \email{urbanow@mpipks-dresden.mpg.de}
 \affiliation{Max Planck
Institute for Physics of Complex Systems\\
 N\"{o}thnitzer Str. 38\\ D--01187 Dresden, Germany}
\author{Janusz A. Ho{\l}yst}%
 \email{jholyst@if.pw.edu.pl}
\affiliation{Faculty of Physics and Center of Excellence for
Complex Systems Research\\Warsaw University of Technology\\
Koszykowa 75, PL--00-662 Warsaw, Poland}
 \affiliation{Max Planck
Institute for Physics of Complex Systems\\
 N\"{o}thnitzer Str. 38\\ D--01187 Dresden, Germany}
\date{\today}

 \begin{abstract}
\par On the basis of a local-projective with nonlinear  constraints (LPNC) approach
(see K. Urbanowicz,  J.A. Ho{\l}yst, T. Stemler  and H. Benner,
Acta Phys. Pol B \textbf{35 (9)}, 2175, 2004) we develop a method
of noise reduction in time series that makes use of constraints
appearing due to the continuous character of flows. As opposed  to
local-projective methods in our method we do not need to determine
the Jacobi matrix.
 The approach has been successfully applied  for
separating a signal from noise in the Lorenz model and in noisy
experimental data obtained from an electronic Chua circuit. The
method was then applied for filtering noise in human voice.
\end{abstract}
\pacs{05.45.Tp,05.40.Ca} \keywords{Chaos, noise reduction, time
series} \maketitle

\section{Introduction}
    \par It is common that observed data  are contaminated
 by noise (for a review of methods of
 nonlinear time series analysis see
 \cite{kantzschreiber,abarbanel,kapitaniak}). The presence of noise can substantially affect
 such system parameters as dimension,
     entropy or Lyapunov exponents \cite{urbanowicz}. In fact noise can completely destroy the fractal
     structure of a chaotic attractor \cite{kostelich} and even $2\%$ of noise can make a dimension
     calculation misleading \cite{Schreiber1}. It follows that both from the theoretical as well as
     from the practical point of view it is desirable to reduce the noise level. Thanks to noise reduction
\cite{kostelich,urbnr03,Schreiber2,Farmer,hammel,davies,zhang,chen,Grassberger,kalman,effern,Hsu,Sauer,Schreiber}
      it is possible e.g. to restore the hidden
    structure of an attractor which is smeared out by noise,
    as well as to improve the quality of predictions.

    \par Every method of noise reduction assumes that it is possible to distinguish between noise
    and a {\it clean signal} on the basis of some objective criteria.
    Conventional methods such as linear filters use a power spectrum for this purpose.
    Low pass filters assume that a clean signal has some typical low frequency,
    respectively it is true for high pass filters.
    It follows that these methods are convenient for a regular source which generates a periodic or
    a quasi-periodic signal.
    In the case of chaotic signals linear filters cannot be used for noise reduction without a substantial disturbance of
    the clean signal. The reason is the broad-band spectrum of chaotic signals. It follows that for chaotic systems
    we make use of another generic feature of dissipative motion that is located on attractors consisting
     of subset of smooth manifolds of an admissible phase space.
    As result corresponding state vectors reconstructed
    from time delay variables are limited to geometric
    objects that can be locally linearized. This fact is a common background of all
    local projective (LP) methods of noise reduction.

    \par Besides the LP approach there are also noise reduction methods
    that approximate an unknown equation of motion and use it
    to find corrections to state vectors. Such methods make use
    of neural networks \cite{zhang} or a genetic programming \cite{chen} and one has to assume
    some basis functions e.g. radial basis functions
    \cite{broomhead} to reconstruct the equation of motion. Another group of
    methods are modified linear filters e.g. the Wiener
    filter \cite{Schreiber}, the Kalman filter \cite{kalman}, or methods based on wavelet analysis \cite{effern}.
    Applications of these methods are limited to systems with large sampling frequencies,
    and they are confined to the locally linear nearest neighborhood of every point in phase space.

    \par The method described in this paper can be considered as an extension
    of a local-projective with nonlinear  constraints (LPNC)
    approach that was introduced in Ref.~\cite{urbnr03}.  We call our method the
    \textit{local projection with nonlinear constraints for flows} (LPNCF).
    The method takes into account natural constraints  that  occur due to  the  continuous
     behavior of flows.

\par The paper is organized as follows.
    In the following section we shall present the LPNCF method and
    the comparison with LP methods is  show in Sec.~\ref{sec.comparison}.
    In Sec.~\ref{sec.examples} we present examples  of noise reduction
    and  application to human voice  filtering.

\section{The LPNCF method}
\par In Ref~\cite{urbnr03} the LPNC method of noise
reduction of deterministic signal is presented. In this paper we
introduce a method that is based on the formulation given in
Ref~\cite{urbnr03} but it brings much better results as compared
to LP approach.
\par Let $\set{x_i}$ for $i=1,2,\ldots,N$ be the time series. The
corresponding clean signal we denote as $\set{\tilde{x}_i}$, so
when the measurement noise $\set{\eta_i}$ is present we come to
the formula $x_i=\tilde{x}_i+\eta_i$ for $i=1,2,\ldots,N$. We can
define the time delay vectors
$\mathbf{x}_i=(x_i,x_{i-\tau},\ldots,x_{i-(d-1)\tau})$ as our
points in the reconstructed phase space. Then we can find two
nearest neighbors $\mathbf{x}_k,\mathbf{x}_j\in\NN$ to vector
$\mathbf{x}_n$ ($\NN$ is the set of nearest neighborhood of the
point $\mathbf{x}_n$). Let us introduce the following function
\cite{urbnr03}
\begin{eqnarray} \mathbf{G}_n(s)=x_{n-s}\lb
x_{k+1-s}-x_{j+1-s}\rb\nonumber\\+x_{k-s} \lb
x_{j+1-s}-x_{n+1-s}\rb+x_{j-s}\lb
x_{n+1-s}-x_{k+1-s}\rb,\label{eq.g}\end{eqnarray} for
$s=0,1,\ldots,d-1$.
%correction 2)
The function $\mathbf{G}_n(s)$ vanishes for clean one-dimensional
    systems because it appears as a constraint after eliminating $a$
    and $b$ from the following equations: \ber
    \tilde{x}_{n+1}=a\tilde{x}_n+b\nonumber\\
    \tilde{x}_{k+1}=a\tilde{x}_k+b\nonumber\\
    \tilde{x}_{j+1}=a\tilde{x}_j+b. \eer In the case of higher
    dimensional systems the function $\mathbf{G}_n(s)$ does not always
    vanish but is altering slowly in time for dense sampling. This is
    because the absolut value of the term $\mathbf{G}_n(s)$ is a function of
    difference of neighboring data $\lb x_{k+1-s}-x_{j+1-s}\rb$ etc.,
    which evolve smoothly in time (near neighbors behave similar in
    consecutive time steps).
%end correction 2)
Now one can check that for a highly sampled clean dynamics there
can be derived such a constraint
\begin{eqnarray} \mathbf{C}_n^m=\sum\limits_{k=0}^{m-1}(-1)^l
\mathbf{G}_n(k)\approx 0,\nonumber\\ (
l=k+\sum_{s=1}^{int(log2(k))}
int(k/2^s))\label{eq.constraint}\end{eqnarray} where $int(z)$ is a
integer part of $z$ and $log2(z)$ is a logarithm with a base $2$
from $z$. For example with $m=8$ the formula~(\ref{eq.constraint})
gives the following
\begin{eqnarray}
\mathbf{C}_n^8=\mathbf{G}_n(0)-\mathbf{G}_n(1)-\mathbf{G}_n(2)+\mathbf{G}_n(3)\nonumber\\-\mathbf{G}_n(4)
+\mathbf{G}_n(5)+\mathbf{G}_n(6)-\mathbf{G}_n(7).\label{eq.8const}\end{eqnarray}
Such a criterium for a constraint can be understood easily if we
notice that all elements $\mathbf{G}_n(s)$ have almost the same
value for clean data and small $s$. Using this we force the second
element of constraint~(\ref{eq.8const}) $\mathbf{G}_n(1)$ to take
the oppose sign as the first element $\mathbf{G}_n(0)$. Then the
group consisting of the third $\mathbf{G}_n(2)$ and fourth
$\mathbf{G}_n(3)$ elements should have the oppose sign to the
group of the first and second element of the
constraint~(\ref{eq.8const}). If we know that elements
$\mathbf{G}_n(s)$ are slightly changing with increasing $s$ the
constraint~(\ref{eq.8const}) should vanish for clean data and for
large enough $m$.

Similarly as in LP methods the constraints~(\ref{eq.constraint})
are ensured in this approach by application of the method of
Lagrange multipliers to an appropriate cost function. Since we
expect that corrections to noisy data should be as small as
possible, the cost function is assumed to be the sum of squared
corrections $S=\sum_{s=1}^N \lb\delta x_s\rb^2$.
\par It follows that we are looking for the minimum of the functional \be
S=\sum_{n=1}^N\lb \delta x_n\rb^2+\sum_{n=1}^N\lambda_n
\mathbf{C}_n^m= min. \label{eq.funkcjonal1}\end{equation} After
finding zero points of $2N$ partial derivatives one gets $2N$
equations with $2N$ unknown variables $\delta x_n$ and
$\lambda_n$. However, in such a case the derivatives of the
functional (\ref{eq.funkcjonal1}) are nonlinear functions of these
variables. For simplicity of computing we are interested to pose
our problem in such a way that linear equations appear which can
be solved by standard matrix algebra. To understand the role of
nonlinearity let us write the terms $\mathbf{G}_n(s)$ in
constraint $\mathbf{C}_n^m$ in such a way that an explicit
dependence on the unknown variables is seen
\begin{eqnarray} \mathbf{G}_n(s)= G\lb
\mathbf{X}_{n-s},\mathbf{X}_{n-s+1}\rb+G\lb \delta
\mathbf{X}_{n-s} ,\mathbf{X}_{n-s+1}\rb+\nonumber\\G\lb
\mathbf{X}_{n-s},\delta\mathbf{X}_{n-s+1}\rb+G\lb \delta
\mathbf{X}_{n-s} ,\delta\mathbf{X}_{n-s+1}\rb.
\label{eq.rozwin}\end{eqnarray} Here we introduced the following
notation
\begin{widetext} \begin{eqnarray} G\lb \mathbf{X}_{n-s},\mathbf{X}_{n-s+1}\rb\equiv x_{n-s}\lb
x_{k-s+1}-x_{j-s+1}\rb+x_{k-s}\lb
x_{j-s+1}-x_{n-s+1}\rb+x_{j-s}\lb
x_{n-s+1}-x_{k-s+1}\rb\nonumber\\
G\lb \delta \mathbf{X}_{n-s} ,\mathbf{X}_{n-s+1}\rb\equiv \delta
x_{n-s}\lb x_{k-s+1}-x_{j-s+1}\rb+\delta x_{k-s}\lb
x_{j-s+1}-x_{n-s+1}\rb+\delta x_{j-s}\lb
x_{n-s+1}-x_{k-s+1}\rb\nonumber\\
G\lb \mathbf{X}_{n-s},\delta\mathbf{X}_{n-s+1}\rb\equiv x_{n-s}\lb
\delta x_{k-s+1}- \delta x_{j-s+1}\rb+x_{k-s}\lb  \delta
x_{j-s+1}- \delta x_{n-s+1}\rb+x_{j-s}\lb
 \delta x_{n-s+1}- \delta x_{k-s+1}\rb\nonumber\\
G\lb \delta \mathbf{X}_{n-s} ,\delta\mathbf{X}_{n-s+1}\rb\equiv
\delta x_{n-s}\lb \delta x_{k-s+1}- \delta x_{j-s+1}\rb+\delta
x_{k-s}\lb \delta x_{j-s+1}- \delta x_{n-s+1}\rb+\delta x_{j-s}\lb
 \delta x_{n-s+1}- \delta x_{k-s+1}\rb,\label{eq.expandg}\end{eqnarray} \end{widetext}

 where
 $\mathbf{X}_{n-s}=\set{x_{n-s},x_{k-s},x_{j-s}}$,
 $\delta \mathbf{X}_{n-s}=\set{\delta x_{n-s},\delta x_{k-s},\delta x_{j-s}}$,
 and $\mathbf{x}_k,\mathbf{x}_j$ are the nearest neighbors of $\mathbf{x}_n$. Indices are defined as
 $\set{n,j,k:\mathbf{x}_n,\mathbf{x}_k,\mathbf{x}_j\in\NN}$. In the case
of uncorrelated noise and under the assumption that the introduced
corrections completely reduce the noise effect $\delta
x_s=-\eta_s\quad(\forall_{s=1,\ldots,N})$ one can neglect the
nonlinear terms in Eqs.~(\ref{eq.expandg}) i.e.
\begin{equation}
 \sum_{s=0}^m G\lb \delta \mathbf{X}_{n-s},\delta
 \mathbf{X}_{n-s+1} \rb \cong 0\quad \forall\; n=1,\ldots,N \label{eq.przybliz}.
\end{equation}
In the equation~(\ref{eq.przybliz}) we use the fact that
$\mean{\eta_i}=0$ and $\mean{\eta_i \eta_j} \sim \delta_{ij}$.
 \par
    Taking into account the assumption (\ref{eq.przybliz}) one can write
    the following \textit{linear equation} for the problem~(\ref{eq.funkcjonal1})
    \be
    \mathbf{M} \cdot \delta
    \mathbf{X}=\mathbf{B},\label{eq.liniowef}
    \end{equation}
    where $\mathbf{M}$ is a matrix containing constant elements,
    $\mathbf{B}$ is a constant vector, and $\delta
    \mathbf{X}^T=(\delta x_1,\delta x_2,\ldots,\delta x_N,
    \lambda_1,\lambda_2,\ldots,\lambda_N)$ are vector dependent
    variables ($T$ - transposition). In practice it is very difficult or even impossible
    to find the solution of the equation~(\ref{eq.liniowef}) for large N.
    First, it is time consuming to solve a linear equation with a matrix $2N\times 2N$ matrix for $N>1000$.
    Second, when $\mathbf{M}$ becomes singular the estimation error of the inverse matrix $M^{-1}$ is very large.
    Third, we cannot always find the true nearest neighbors (the set $\NN$ for clean dynamics) from
    the noisy data $\set{x_i}$. Taking into account the above reasons it is useful to replace
     the global minimization problem (\ref{eq.funkcjonal1}) by $N$ local minimization problems
     related to the nearest neighborhood
     $\NN $. The corresponding local functionals to be minimized are\begin{eqnarray}
     S_n^{NN}=\sum_s \lb \delta x_s
    \rb^2+\lambda_n C_n^m = min\nonumber\\ \forall\; n=1,...,N\quad\mbox{where} \quad\set{s:x_s\in \NN\;\mbox{or}\;x_s\in
    \mathbf{X}_{n+1}}.\label{eq.minimalizlok}
    \end{eqnarray}
    We can consider the minimization problem~(\ref{eq.minimalizlok}) as a certain
approximation of (\ref{eq.funkcjonal1}). The global
problem~(\ref{eq.funkcjonal1}) is equivalent to
Eq.~(\ref{eq.liniowef}) with $2N$ unknown variables that should be
found single-time. The problem (\ref{eq.minimalizlok}) is
equivalent to a system of coupled equations that should be solved
several times and as a result one gets an approximate global
solution. Writing Eq.~(\ref{eq.minimalizlok}) in the linear form
i.e. calculating the zeros of corresponding derivatives and using
Eq.~(\ref{eq.przybliz}) one gets $N$ linear equations as follows
\be  \mathbf{M}_n \cdot \delta
\mathbf{X}_n^{\lambda}=\mathbf{B}_n\quad \forall\;
n=1,\ldots,N,\label{eq.liniowezlok}\end{equation} where $\lb\delta
\mathbf{X}_n^{\lambda}\rb^T=(\delta x_n,\delta x_k,\delta
x_j,\delta x_{n+1},\delta x_{k+1},\delta x_{j+1},\lambda_n)$. The
matrices $\mathbf{M}_n$ corresponding to (\ref{eq.minimalizlok})
avoid the disadvantages of (\ref{eq.liniowef}), i.e. they are not
singular, their dimension is small and they do not substantially
depend on the initial approximation of nearest neighbors. Matrix
$\mathbf{M}_n$ for $m=1$ is given by
\begin{widetext}
\begin{eqnarray} \mathbf{M}_n=\left[
  \begin{array}{ccccccc}
    2 & 0 & 0 & 0 & 0 & 0 & x_{k+1}-x_{j+1}\\
    0 & 2 & 0 & 0 & 0 & 0 & x_{j+1}-x_{n+1}\\
    0 & 0 & 2 & 0 & 0 & 0 & x_{n+1}-x_{k+1}\\
    0 & 0 & 0 & 2 & 0 & 0 & x_j-x_k\\
    0 & 0 & 0 & 0 & 2 & 0 & x_n-x_j\\
    0 & 0 & 0 & 0 & 0 & 2 & x_k-x_n\\
    x_{k+1}-x_{j+1} & x_{j+1}-x_{n+1} & x_{n+1}-x_{k+1} & x_j-x_k & x_n-x_j & x_k-x_n & 0
  \end{array}
\right]
 \end{eqnarray}\end{widetext}
   Vector $\mathbf{B}_n$ has the form
    $\mathbf{B}_n^T=\set{0,0,0,0,0,0,-\mathbf{G}_n(0)}$. Note that
    this matrix in one-dimensional case is the same for LPNC
    method given in Ref.~\cite{urbnr03}, but constraints and matrix $\mathbf{M}_n$
    will essentially differ in higher dimensions for both methods.
 \section{\label{sec.comparison}Comparison to standard LP methods}
\par
\begin{figure}
\includegraphics[scale=0.6]{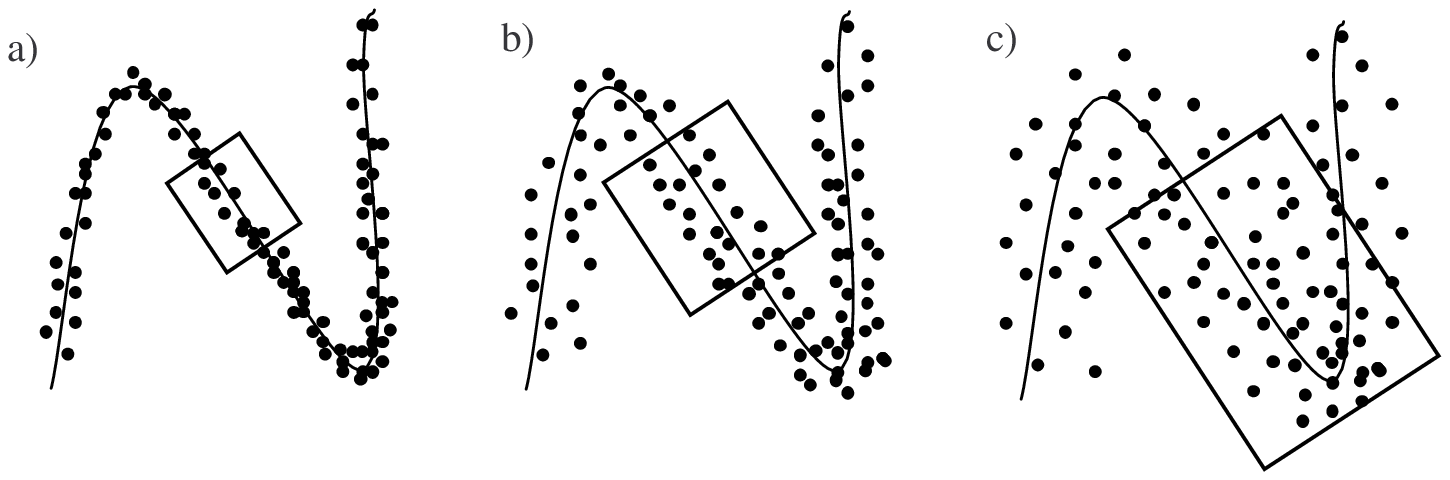}
\caption{\label{fig.2} The plot of the clean attractor (continuous
line), noisy data connected to this attractor (black dots) as well
as range of the nearest neighborhood taken under consideration to
determine tangent subspace (rectangle). The level of noise for the
case c) is so high that a linear approximation is not longer
valid.}
\end{figure}

 Minimizations problems used in standard LP methods and in this method
 Eq.~(\ref{eq.constraint}) are not equivalent because in our  case  we do not have to estimate the Jacobi matrix
 at all.
 These differences in practice are as
 follows a) Eq.~(\ref{eq.constraint}) is nonlinear against corrections
$\delta x_i$.  The approximation in this case means a
corresponding linearization. b)  For constraints in standard LP
methods we do not know the exact values of Jacobi matrix
$\mathbf{A}$. The approximation means that Jacobi matrix
$\mathbf{A}$ is estimated from noisy data. The LP methods look for
subsequent corrections to noisy data by finding of a subspace
tangent to an unknown attractor corresponding to the clean
dynamics and projecting noisy data on this subspace. If one tries
to estimate the position of the tangent subspace, what is
equivalent to estimation of the Jacobi matrix $\mathbf{A}$ from
noisy data, the range of the neighborhood should be larger than
the magnitude of noise. Such a procedure should allow to
distinguish between the dominating direction (connected with
system dynamics) and random directions connected with a noise (see
Figs.~\ref{fig.2}a and \ref{fig.2}b). If the noise level is very
high it is not possible to use the tangent subspace to find the
attractor of the clean dynamics since the range of the necessary
nearest neighborhood would be very large
  and the linear approximation
would be invalid (see Fig.~\ref{fig.2}c) \cite{kantzholyst}.
On the other hand, if we consider the minimalization
problem~(\ref{eq.funkcjonal1}) we do not need to find the Jacobi
matrix $\mathbf{A}$ but only to take into account the constraint
equation (\ref{eq.constraint}). Such an approach makes it possible
to use a neighborhood smaller than the noise magnitude and in our
approach the corresponding number of nearest neighbors equals to
$2$. Note that here we encounter a flow, so nearest neighbors
searching is not so biased as local projection.
 %One has to stress that the correction following from a single
%constraint does not correspond to the exact solution and only a
%system of constraints converges  to the clean attractor.
To find two nearest neighbors $\NN$ to $\mathbf{x}_n$ we use  the
Delaunay triangulation \cite{allie} and  the  method to find is
given  in Ref.~\cite{urbnr03}. Searching nearest neighbors by
Delaunay triangulation is very time consuming. That is why we
first look for the $N_{nn}$ nearest  neighbors by means of
Euclidian distance minimization and then perform the Delaunay
triangulation only on this nearest neighborhood. This approach is
as fast as standard nearest  neighbors searching. Accordingly to
the needs, e.g. in online noise reduction  of  human voice, one
can think about speeding up the  method. As we mention in our
approach we need only two nearest neighbors as opposed  to
standard  LP methods, so looking only  for  the two nearest
neighbors close in time would make the searching for closest
neighbors very fast and the method robust against high
non-stationarity. \par One can ask on the smallest sampling rate
per cycle $\mathcal{RS}$ which makes the method applicable. As we
have checked the method works even for the  rate $\mathcal{RS}$
comparable with the $m$ parameter  used in
Eq.~(\ref{eq.constraint}) but  then the efficiency  is smaller
than for the standard LP method. The method works the best for
more than $2\cdot m$ samples per cycle. We have taken the
embedding window $d\cdot\tau$ used in nearest neighbor searching
as long as one cycle. The method is robust against changing the
number of taken nearest neighbors, as opposed to standard LP
methods what will be seen in the next section.

\section{\label{sec.examples}Noise reduction: examples and applications}
\par
\begin{figure}
    \includegraphics[angle=-90,scale=0.35]{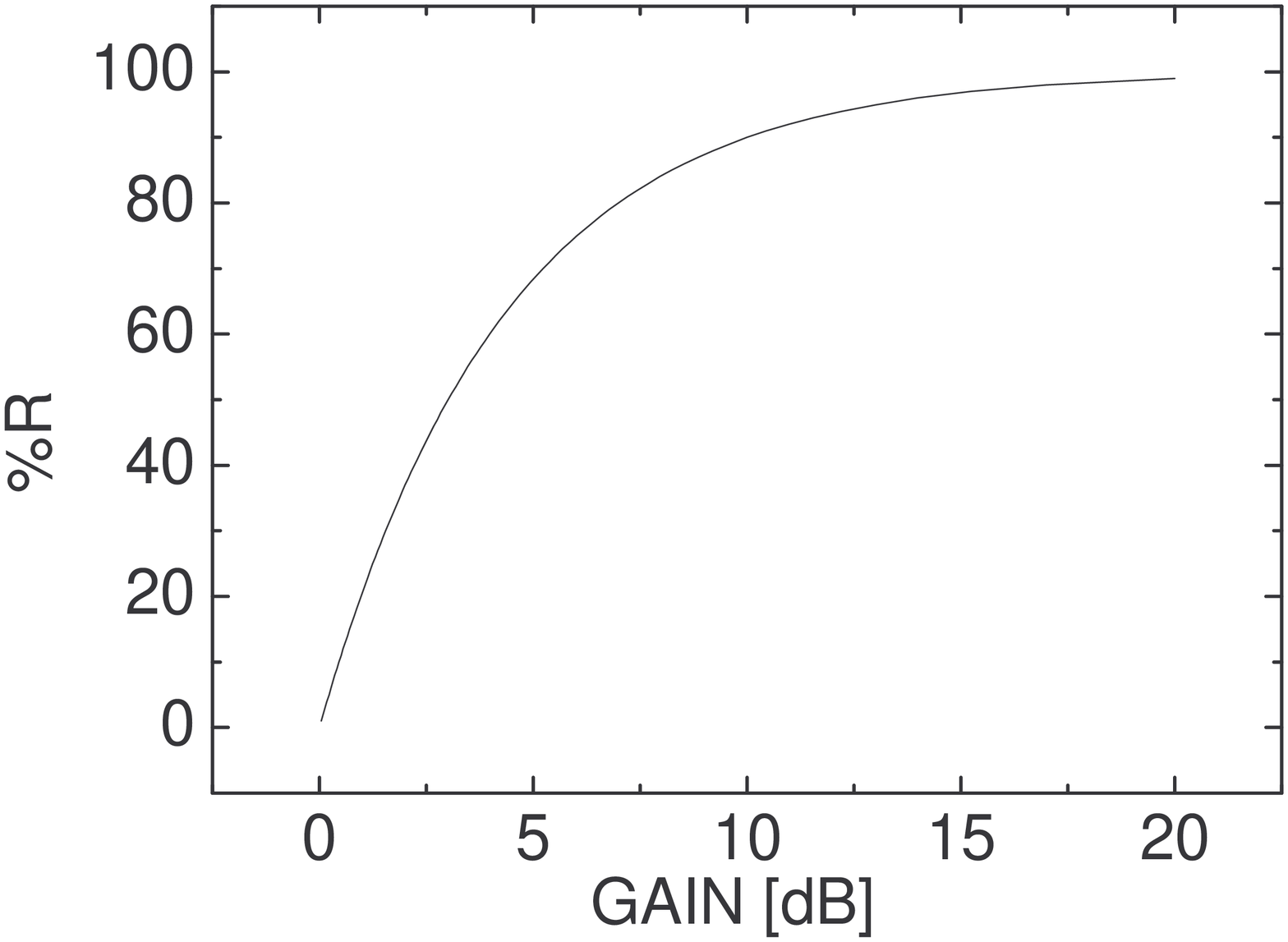}
    \caption{\label{fig.Rgain} The plot of the $\%R$ parameter on the
    gain parameter $\mathcal{G}$.}
    \end{figure}
\begin{figure}
\includegraphics[angle=-90,scale=0.35]{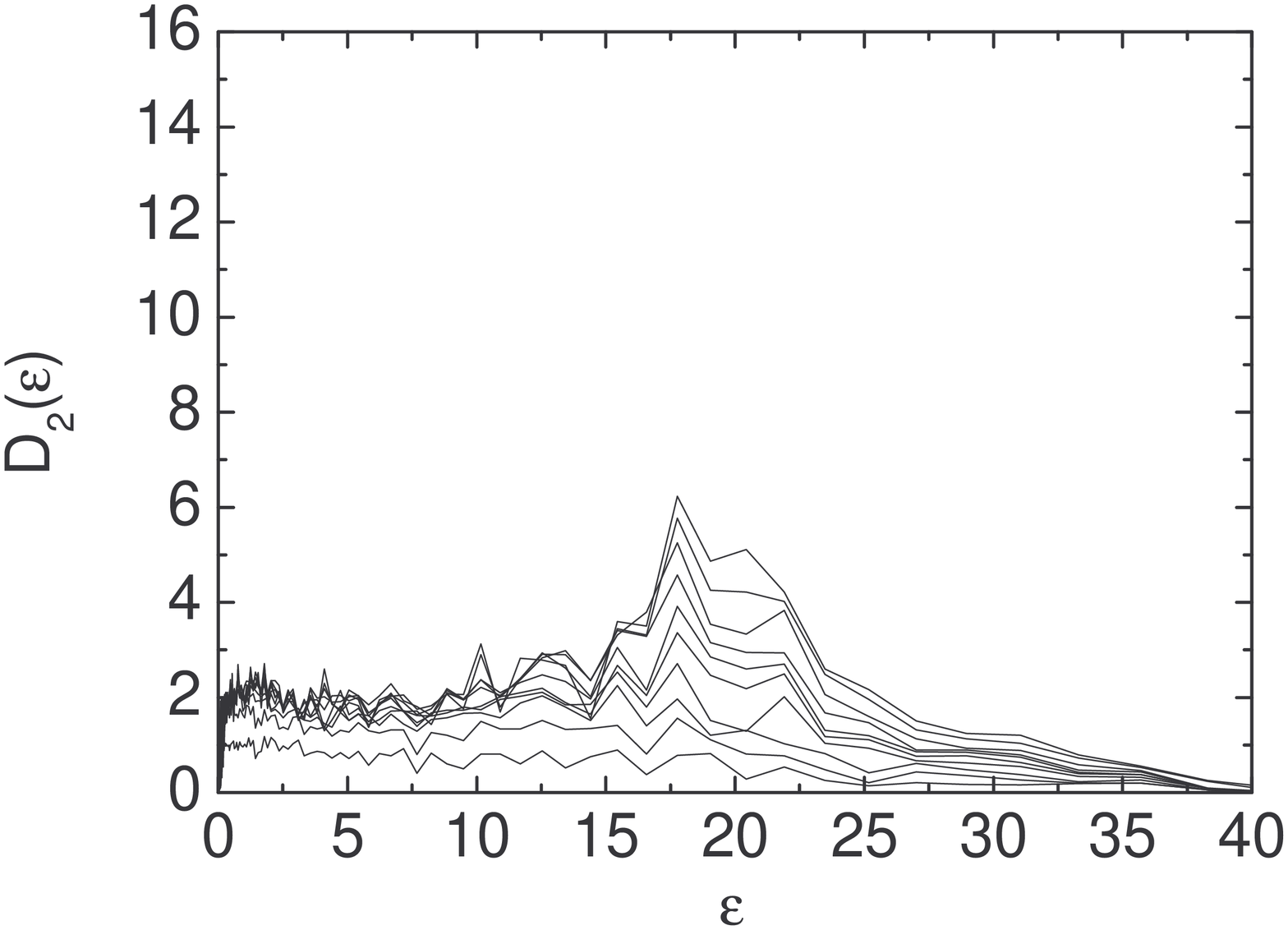}
\caption{\label{fig.d2clean} The plot of the correlation dimension
$D_2$ versus the threshold $\eps$ for the clean Lorenz system.}
\end{figure}
\begin{figure}
\includegraphics[angle=-90,scale=0.35]{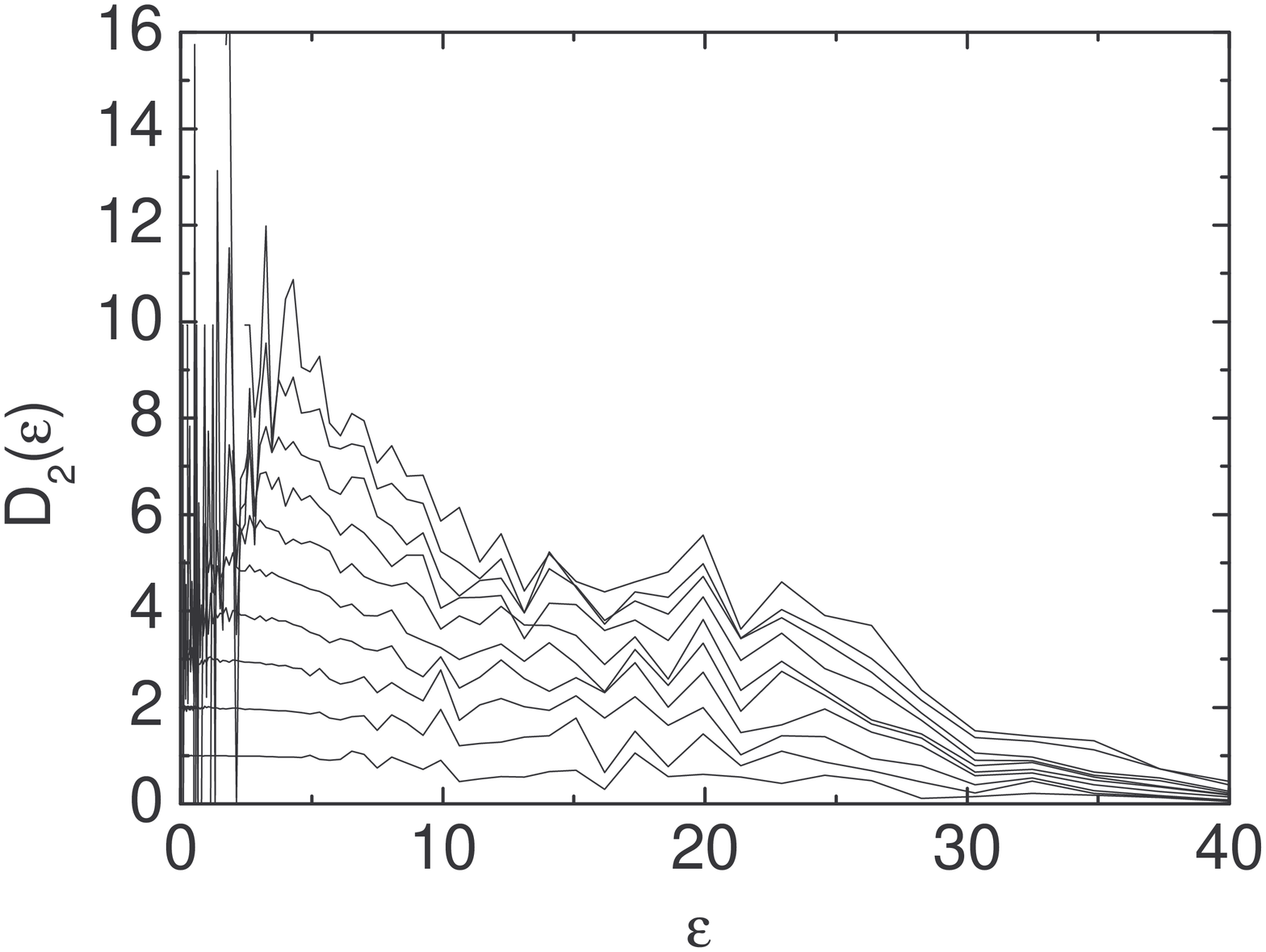}
\caption{\label{fig.d2noise} The plot of the correlation dimension
$D_2$ versus the threshold $\eps$ for the Lorenz system with noise
$\mathcal{N}=48\%$.}
\end{figure}
\begin{figure}
\includegraphics[angle=-90,scale=0.35]{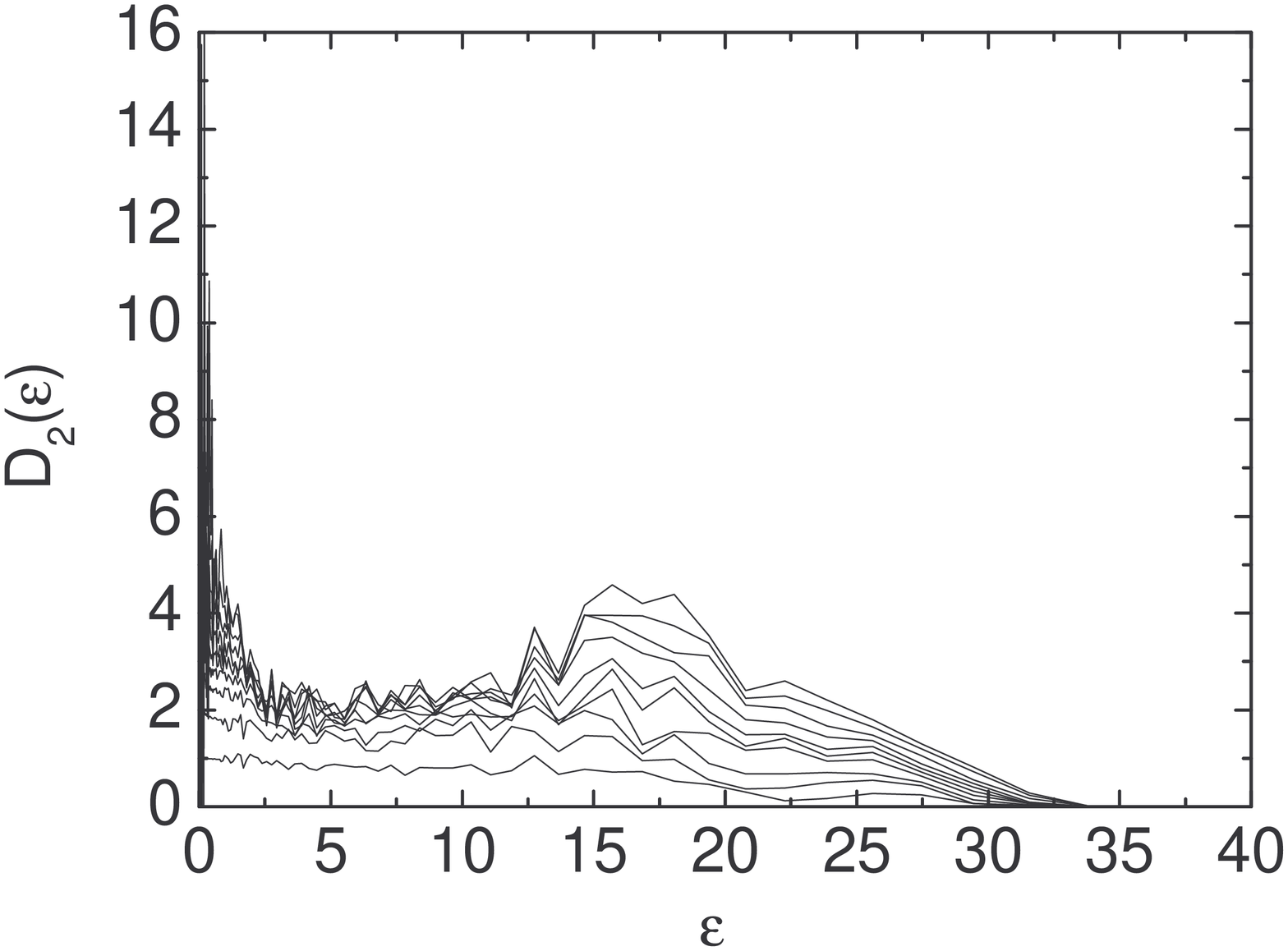}
\caption{\label{fig.d2cl} The plot of the correlation dimension
$D_2$ versus the threshold $\eps$ for the Lorenz system with noise
after noise reduction.}
\end{figure}

Let us define the noise level $\mathcal{N}$ \be
\mathcal{N}=\frac{\sigma}{\sigma_{DATA}}, \ee where $\sigma$ is a
standard  deviation of noise and $\sigma_{DATA}$ is the standard
deviation of data. The efficiency of noise reduction we  calculate
by  means of the gain parameter which is defined as \be
\mathcal{G}=10 \log\lb
\frac{\sigma_{noise}^2}{\sigma_{red}^2}\rb\end{equation} where
$\sigma_{noise}^2$ is the variance of added noise and
$\sigma_{red}^2$ is the variance of noise left after noise
reduction.
% correction 4a)
The gain parameter can be transformed into another parameter:
$\%R$,
    which says how much noise is reduced:
    %$\%R=(1-10^{-\mathcal{G}/10})*100\%$
    $\%R=(1-\sigma^2_{red}/\sigma^2_{noise})\cdot 100\%$. In
    Fig.~\ref{fig.Rgain} the dependence of the $\%R$ parameter on the
    basic parameter $\mathcal{G}$ is presented. The gain parameter
    $\mathcal{G}$ is commonly used in the evaluation
     of the noise reduction because it gives more relevant information especially in the regime $\%R>90\%$.
% end correction
%correction 4b)
We use the standard Lorenz model to evaluate the performance of
noise reduction methods.
    The Lorenz system is described by a system of three coupled
    differential equations \ber
      \dot{x}=\xi\cdot (y-x)\nonumber\\
      \dot{y}=\rho x-y-xz\label{lorenz_eq}\\
      \dot{z}=xy-\beta z.\nonumber
    \eer The model is widely used for a description of Rayleigh-Benard
    instabilities in fluids \cite{lorenz} and in quantum optics
    for laser dynamics \cite{lorenzoptics}. We use standard parameters
    for this system, i.e. $\xi=10;\rho=28;\beta=8/3$, for which the
    standard "butterfly" attractor can be observed.
% end correction
To verify our method in a real experiment we have performed
analysis of data generated by a nonlinear electronic circuit.
%correction 4a2)
The Chua circuit is one of the simplest electronic nonlinear
    system that exhibits chaotic behaviour \cite{chua1,chua2}.
    The nonlinearity comes from two paralel connected negative
    resistors which are realized by amplifiers with a corresponding feedback.
    The Chua circuit has been studied in the presence of noise added to the outcoming signal.
    The noise (white and Gaussian) has been generated by an
    electronic noise generator.
% end correction 4a2)
%correction 1)
The LPNCF scheme of noise reduction  improves estimations of
    invariant parameters.  Figs~\ref{fig.d2clean}-\ref{fig.d2cl}
    present calculations of the correlation dimension $D_2$ versus
    threshold $\eps$ for the clean Lorenz system, the Lorenz system
    with noise and the latter parameter after noise reduction
    respectively. Using a standard procedure one looks for a plateau
    in an intermediate threshold region. In fact one can observe that
    the plateau $D_2\approx 2$ cannot be found  at
    Fig.~\ref{fig.d2noise}  corresponding to the noisy Lorenz system
    with the noise level $\mathcal{N}=48\%$ but is well seen  after
    the noise reduction with the LPNCF method (see
    Fig.~\ref{fig.d2cl}).

% end correction 1)
\par We have made a quantitative comparison between LPNCF and GHKSS
method. GHKSS method \cite{Schreiber} is the implementation of the
standard LP approach that we think are optimal and most efficient.
In case of both methods we use the same scheme of neighbors
searching, i.e. the minimization of Eucleadian distance, in
addition for LPNCF method we apply then the Delaunay search which
is not needed for GHKSS method. For large sampling rate we did
always some time averaging with curvature correction
\cite{Schreiber} that improve the gain in both methods. For all
the calculations we use $8$ iterations of both methods. The
projection dimension for GHKSS $3-12$ and $m$ for LPNCF method are
in the range $4-64$.
%correction 3)
We can say that the complexity in programming of the both methods
    is comparable. The LPNCF method implemented at 2.5GHz computer  is
    approximately 2-3 times slower to be used for  on-line noise
    reduction in voice and to receive a substantial improvement of the
    voice recognition.
%end correction 3)

\begin{figure}
\includegraphics[angle=-90,scale=0.35]{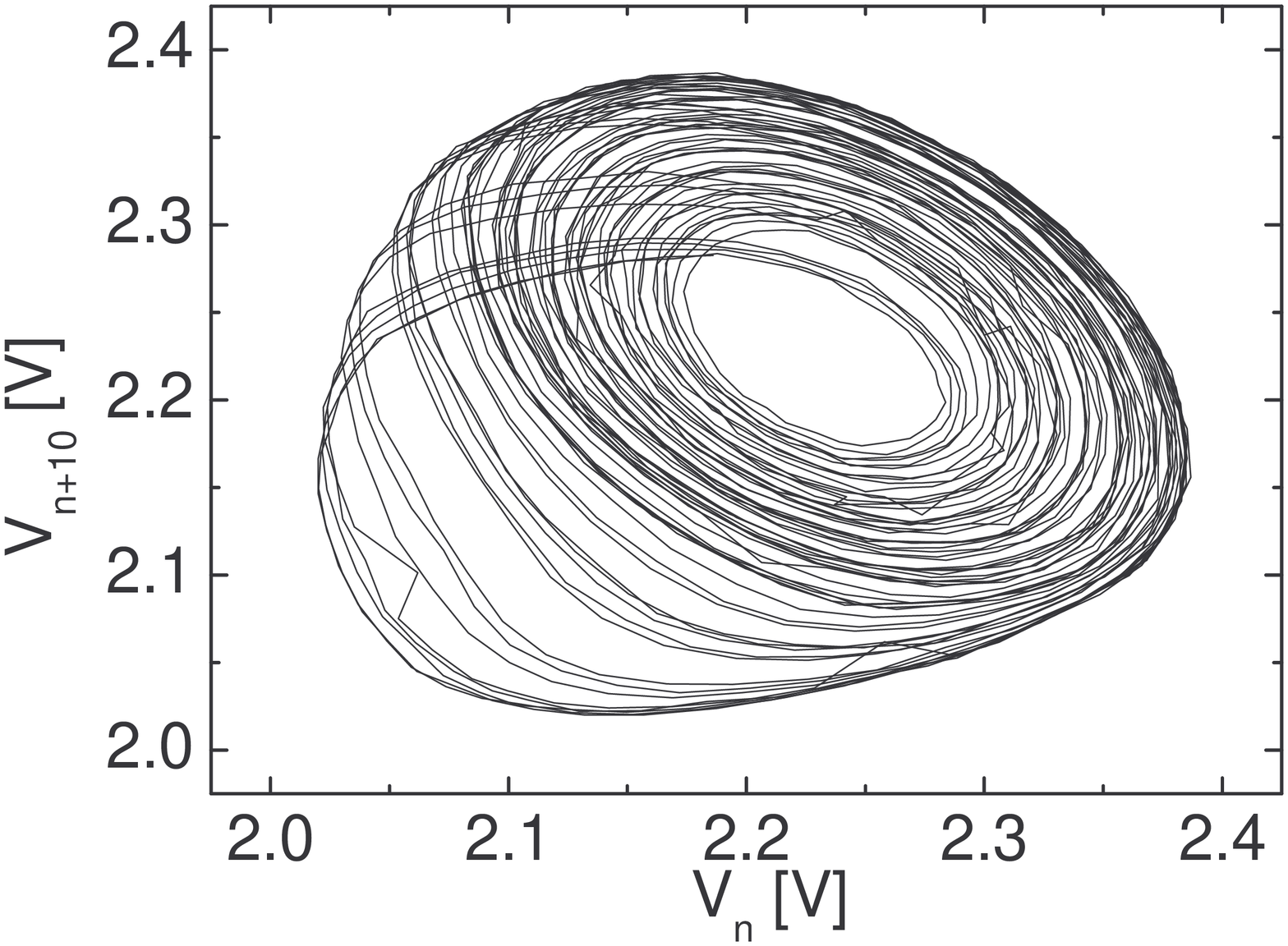}
\caption{\label{fig.chua7000} The sampling corresponding to a
clean trajectory in the Chua circuit (real experiment).}
\end{figure}

\begin{figure}
\includegraphics[angle=-90,scale=0.35]{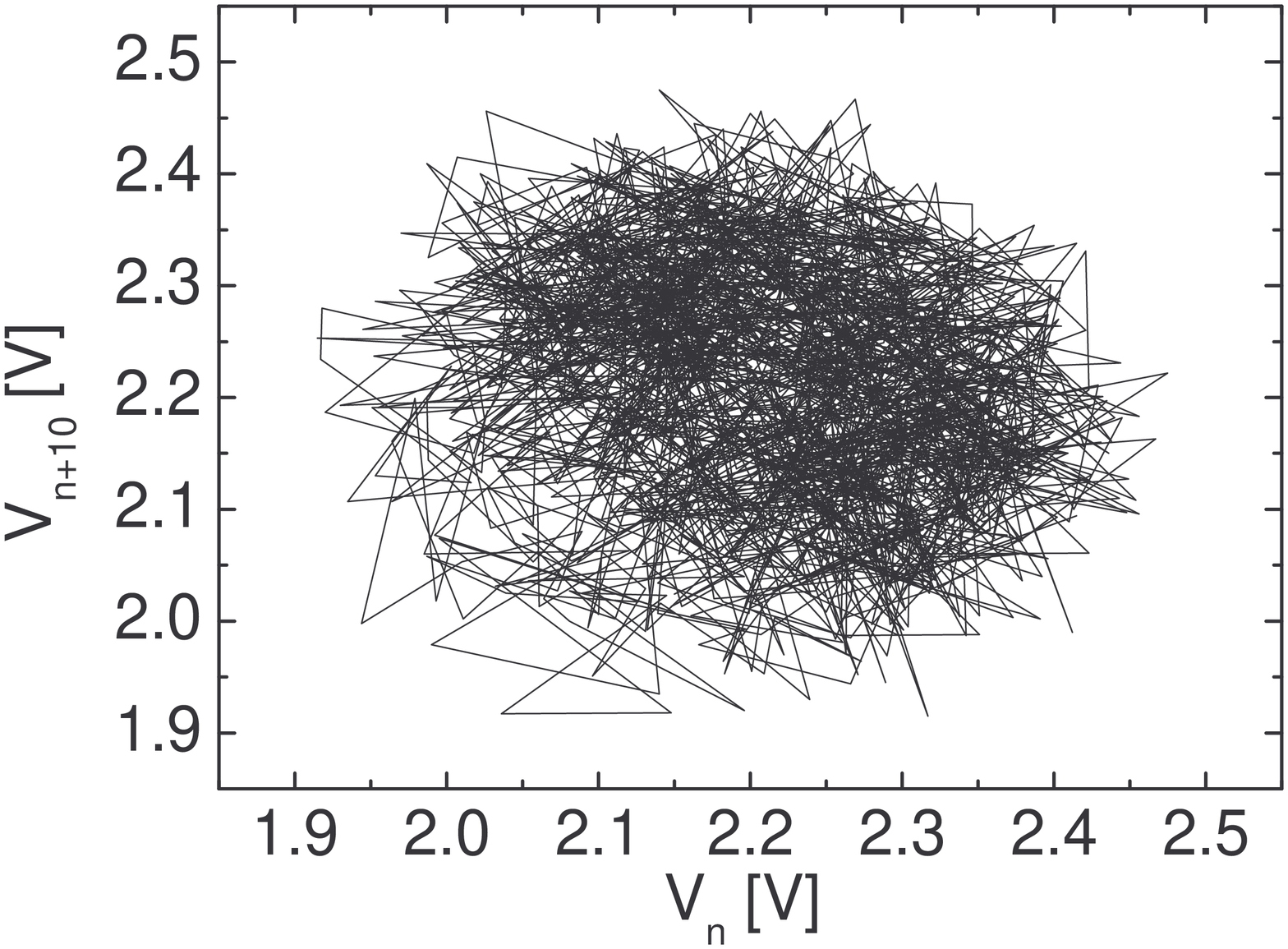}
\caption{\label{fig.chua7000no} The sampling received from the
Chua circuit in the presence of a measurement noise
$\mathcal{N}=46\%$. Note the difference in scale.}
\end{figure}

\begin{figure}
\includegraphics[angle=-90,scale=0.35]{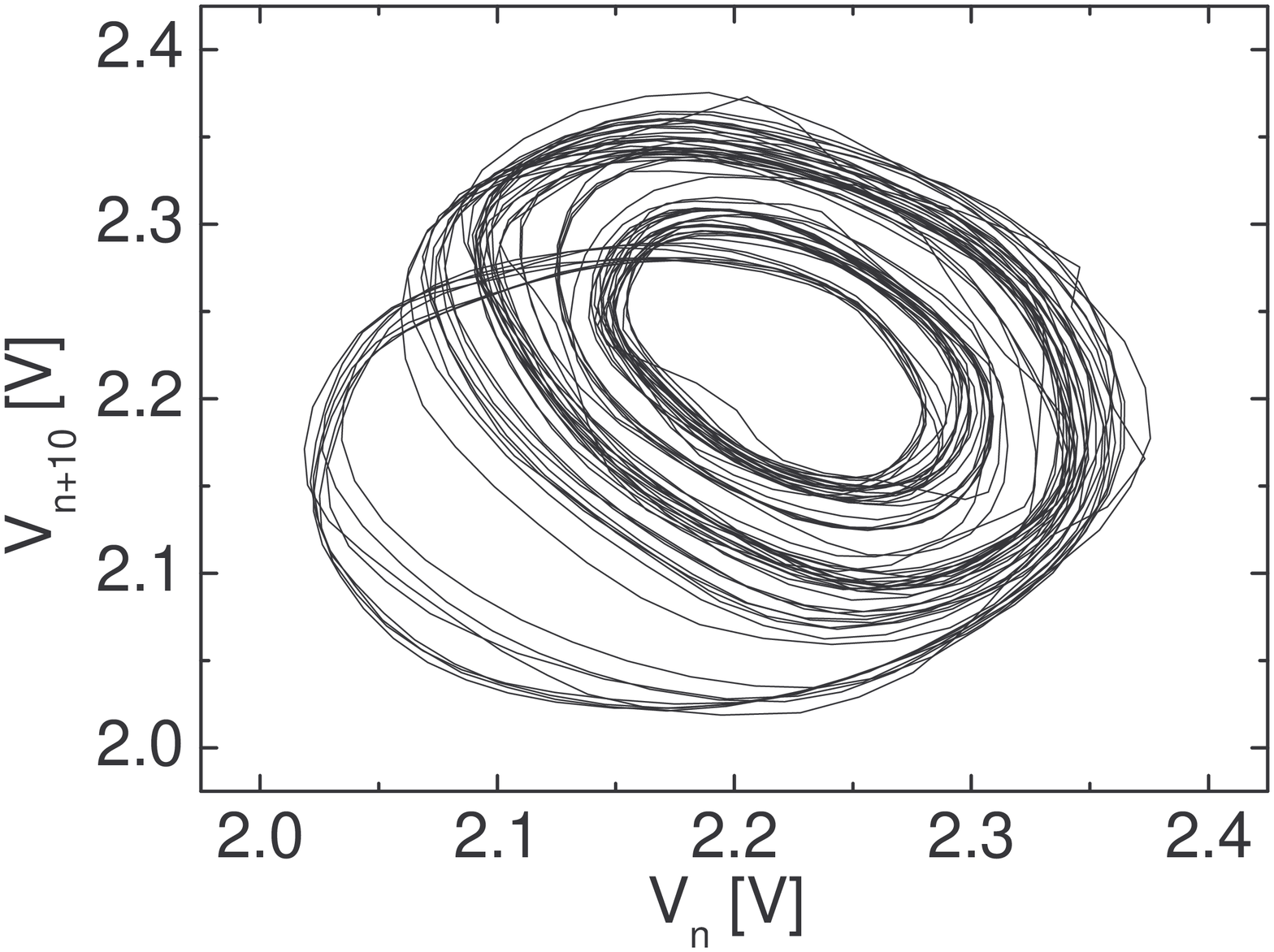}
\caption{\label{fig.chua7000cl} The sampling of Chua circuit
received after the noise reduction applied to data presented at
Fig.~\ref{fig.chua7000no}.}
\end{figure}

Figs.~\ref{fig.chua7000}-\ref{fig.chua7000cl} shows the clean Chua
attractor, with  measurement noise $\mathcal{N}=46\%$ and  after
noise reduction respectively. It is used the LPNCF approach for
$m=3-12$ when sampling ratio was about $50$ per full cycle. The
efficiency was $\mathcal{G}=9.48$ ($\%R=88.7\%$) when the GHKSS
method did $\mathcal{G}=9.33$ ($\%R=88.3\%$).

\begin{figure}
\includegraphics[angle=-90,scale=0.35]{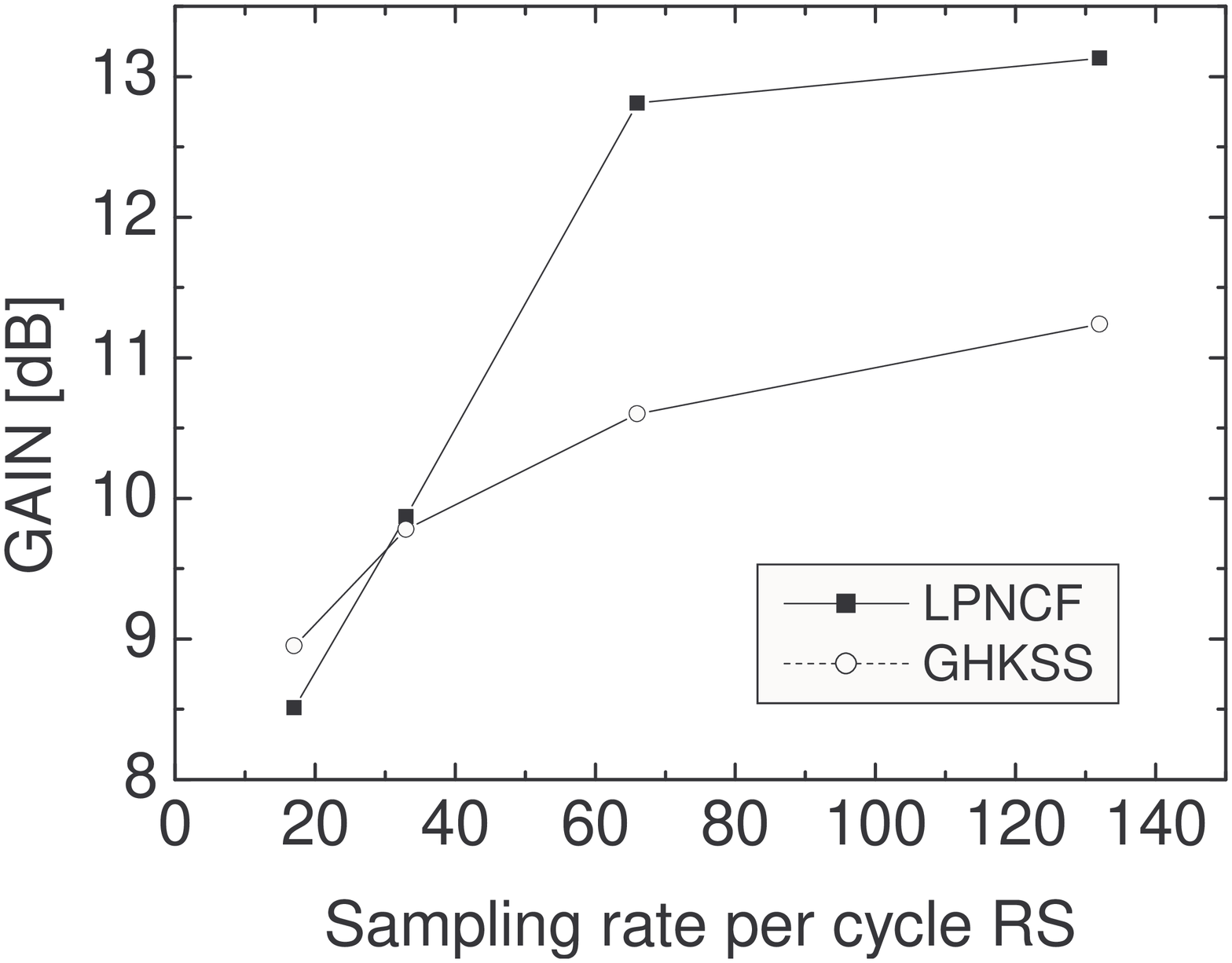}
\caption{\label{fig.SRPC} The efficiency of noise reduction by
LPNCF and GHKSS method for different sampling rate $\mathcal{RS}$.
Here the Lorenz system \cite{lorenz} was  used
($\mathcal{N}=48\%$, $N=5000$, $N_{nn}=20$).}
\end{figure}

\begin{figure}
\includegraphics[angle=-90,scale=0.35]{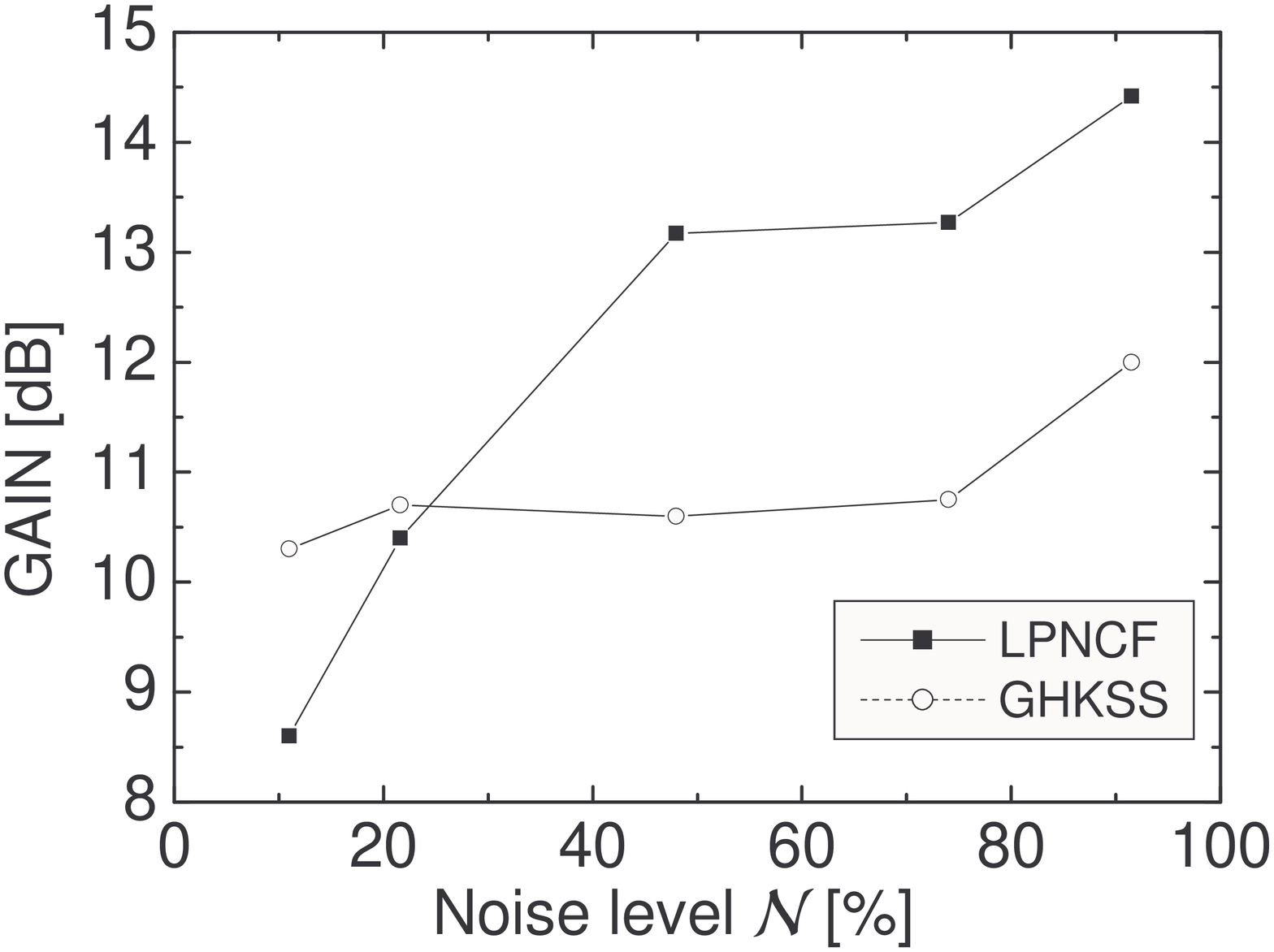}
\caption{\label{fig.NTS} The efficiency  of noise reduction by the
LPNCF and GHKSS method for different noise level $\mathcal{N}$.
Here the Lorenz system \cite{lorenz} was used ($\mathcal{RS}=66$,
$N=5000$, $N_{nn}=20$).}
\end{figure}

\begin{figure}
\includegraphics[angle=-90,scale=0.35]{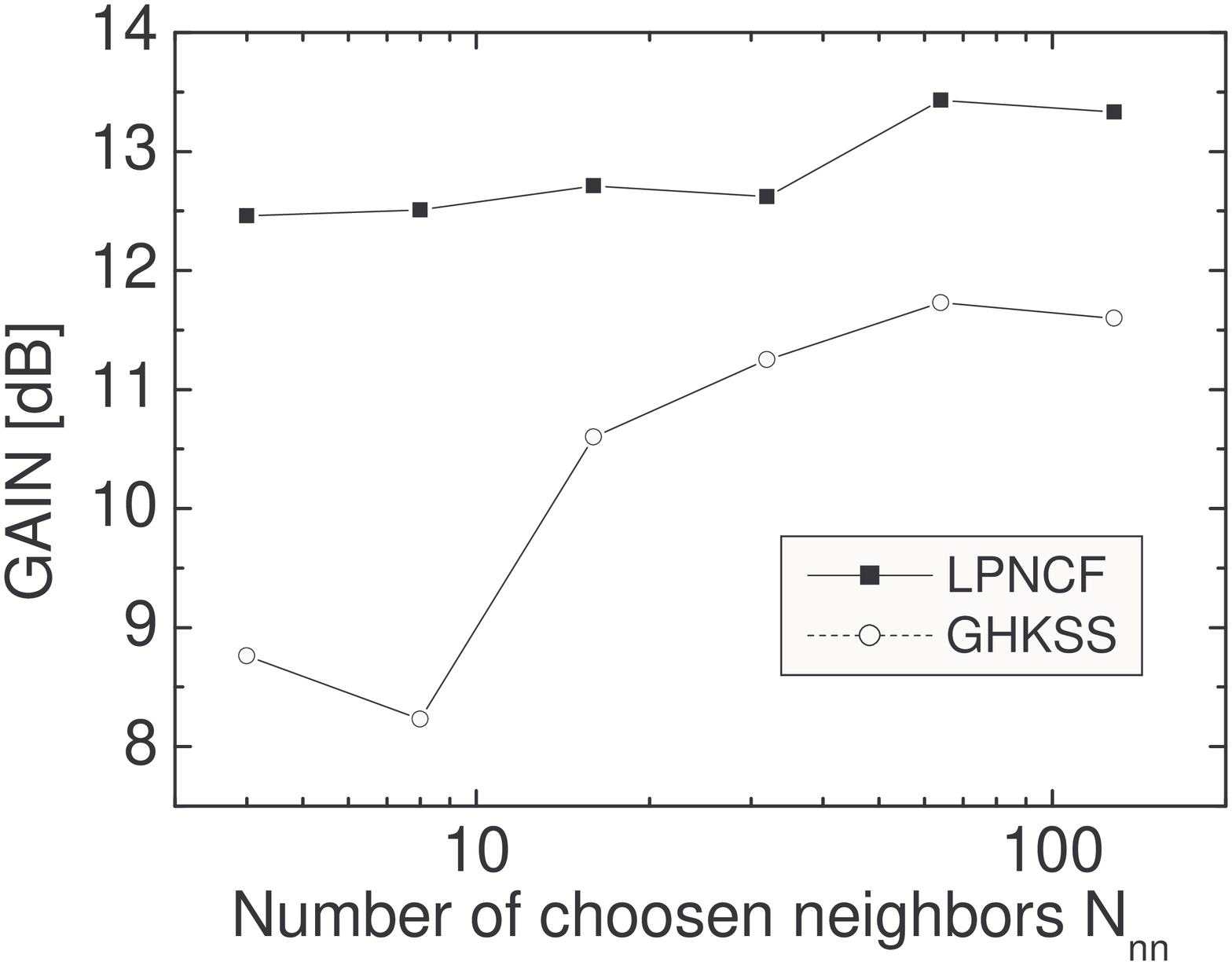}
\caption{\label{fig.Nnn} The efficiency  of noise reduction by the
LPNCF and GHKSS method for various number of neighbors $N_{nn}$.
Here the Lorenz system \cite{lorenz} was used ($\mathcal{N}=48\%$,
$\mathcal{RS}=66$, $N=5000$).}
\end{figure}

We analyze the behavior of noise reduction by the LPNCF and the
GHKSS method against increasing sampling rate per cycle
$\mathcal{RS}$ (see Fig.~\ref{fig.SRPC}). It is clear that the
efficiency of LPNCF method should increase for larger sampling
rate $\mathcal{RS}$. In the figure one can see for large
$\mathcal{RS}$ that the LPNCF method is more efficient than GHKSS
method while it is less efficient for small $\mathcal{RS}$. We
compare the efficiency of these two methods for various noise
levels $\mathcal{N}$ (see Fig.~\ref{fig.NTS}). One can see that
LPNCF method in this case is more efficient for high noise levels
starting from $30\%$. This is because for large noise levels it is
difficult in GHKSS method to determine properly the tangent
subspace as it was explained in the previous section. The last
comparison of these two methods shows the dependence on the number
of regarded neighbors $N_{nn}$ (see Fig.~\ref{fig.Nnn}). In the
LPNCF method the parameter $N_{nn}$ describes the number of
neighbors that are used in preliminary search for candidates to
the Delaunay procedure. It is shown in this figure that for small
number of neighbors the LPNCF method can be used without the loss
of efficiency. Such a behavior is very useful for non-stationary
data, when the large number of neighbors could not be found or
when because of correlated noise one should omit neighbors close
in time. The GHKSS method did some noise reduction for small
$N_{nn}$ because here most corrections come from the time
averaging which alone made $\mathcal{G}=7.2$ ($\%R=80.9\%$).

\par
\begin{figure}
\includegraphics[angle=-90,scale=0.35]{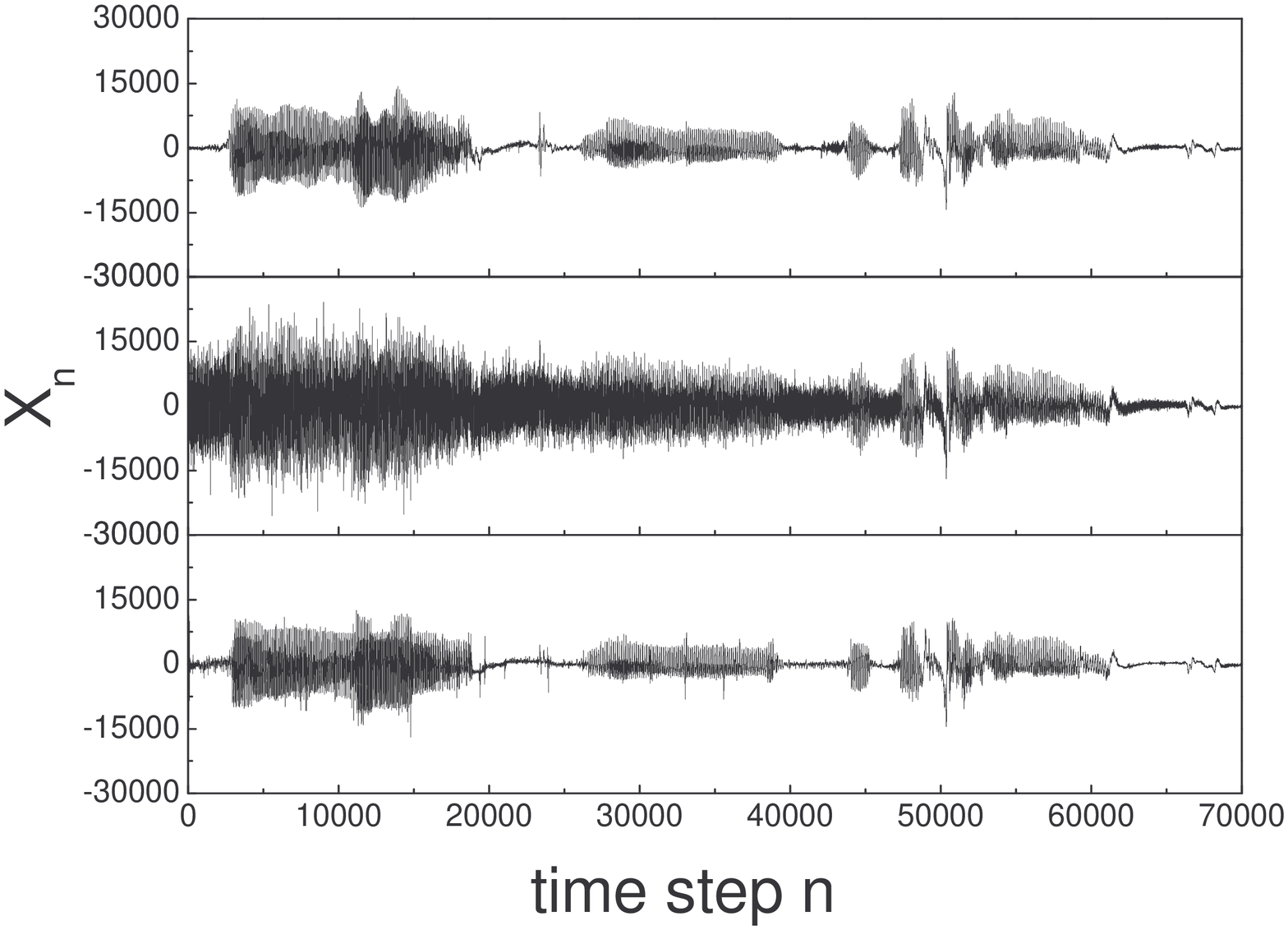}
\caption{\label{fig.tshello} The voice time series of sentence
"Hello world, my name is Krzysztof Urbanowicz". From the upper
panel to the bottom there are clean signal then a series with
decreasing measurement noise and a noisy signal after noise
reduction respectively.}
\end{figure}

\begin{figure}
\includegraphics[angle=-90,scale=0.35]{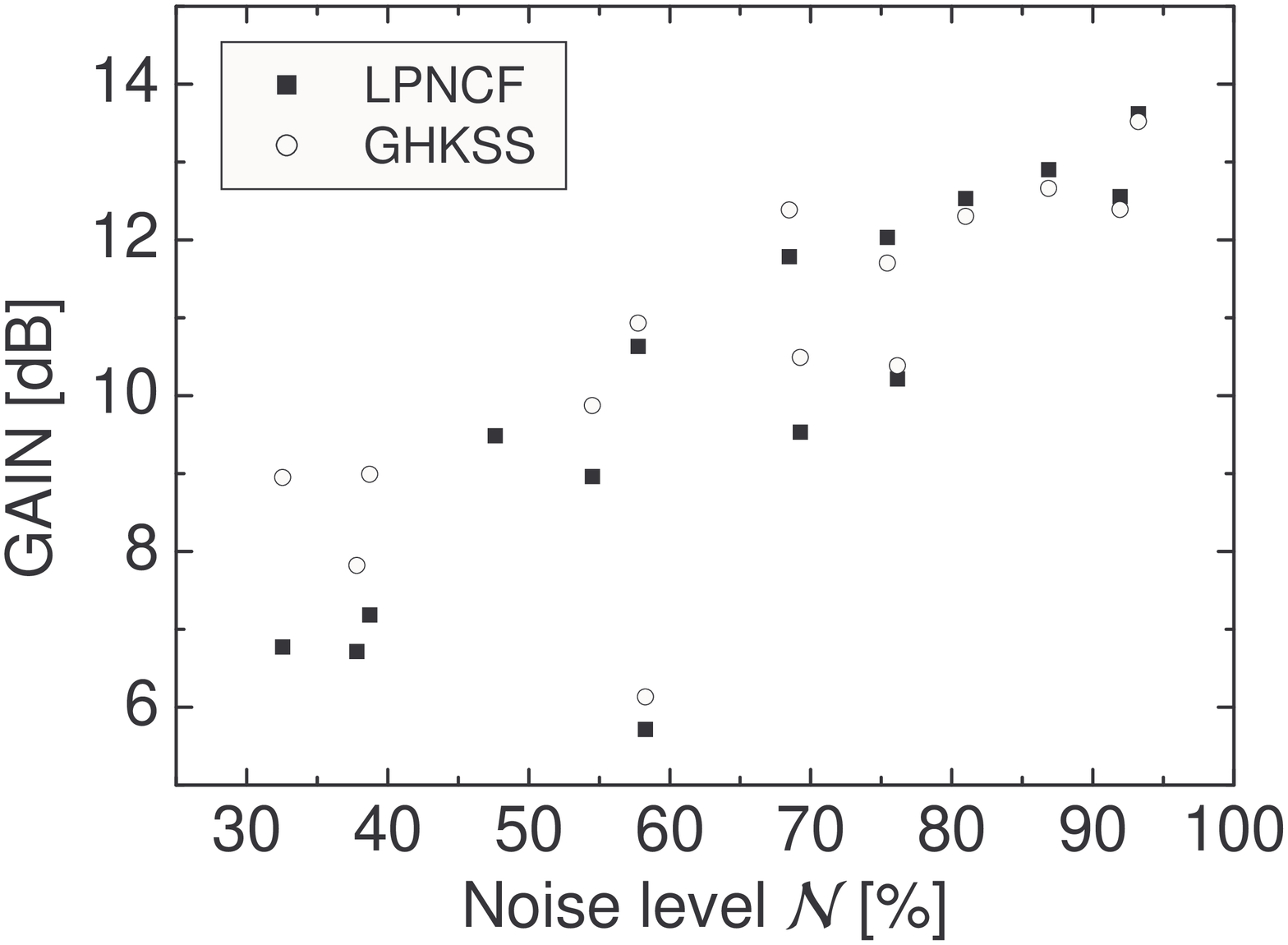}
\caption{\label{fig.helloNTS} The efficiency of noise reduction of
human voice by the LPNCF and GHKSS method for different noise
level $\mathcal{N}$ ($N=5000$).}
\end{figure}

 We applied successfully our method to noise reduction from human
voice \cite{kantzhuman}. On Fig.~\ref{fig.tshello} we show a clean
time series of the recorded sentence "Hello world, my name is
Krzysztof Urbanowicz" (upper panel), this time series with
temporally decreasing measurement noise (middle panel) and after
noise reduction (bottom panel). Noise reduction was made in
windows of length $N=5000$, with parameter $m=3-12$. The voice was
recorded with sampling $22050Hz$ what gives $\mathcal{RS}\approx
120$. The embedding window $d\cdot\tau=100$, so it covers almost
the whole cycle. Fig.~\ref{fig.helloNTS} presents the efficiency
of LPNCF and GHKSS methods. As it is suspected the LPNCF method
did less for small noise levels (see for the comparison
Fig.~\ref{fig.NTS}). Note that here we use larger number of
neighbors than at Fig.~\ref{fig.NTS} i.e. $N_{nn}=60$. For large
noise level both methods work comparably.
% correction 4c)
The gain of the noise reduction for whole data set shown in
Fig.~\ref{fig.helloNTS} is $\mathcal{G}=11.4$ ($\%R=92.8\%$) for
LPNCF and $\mathcal{G}=11.7$ ($\%R=93.2\%$) for GHKSS. Such values
of noise reduction improve significantly voice recognition for
intermediate noise levels. After the performed noise reduction the
background noise is not heard in the recorded signal.

%end correction

\section{Conclusions}
\par In conclusion we developed the method of noise reduction
design for flows. It uses a nonlinear constraints that appear due
to the continuous behavior of flows. To efficiently perform the
noise reduction one needs to find only two nearest neighbors. The
method is robust against input parameters estimation as well as
for highly non-stationarity data. We applied with success the
method for noise from human voice separating.

\begin{acknowledgments}
We acknowledge Holger Kantz for fruitful discussions. The work of
JAH was supported by a grant {\it Dynamics of Complex Systems} of
Warsaw University of Technology and by the COST Action P10 {\it
Physics of Risk}.
\end{acknowledgments}

\newpage
%\bibliography{apssamp}

\end{document}